\documentclass[aps, twocolumn, superscriptaddress, pre, reprint, floatfix, numbers, sort, compress]{revtex4-2}

\usepackage{graphicx}
\usepackage{amssymb,amsmath,amsbsy,amsfonts,bm}
\usepackage[colorlinks=true,linkcolor=black,citecolor=black,urlcolor=black]{hyperref}
\usepackage{epsfig}
\usepackage{multirow}
\usepackage{tikz}
\usetikzlibrary{calc}
\usepackage{threeparttable}
\usepackage{adjustbox}
\usepackage{booktabs}
\usepackage{diagbox}
\usepackage{array}
\usepackage{lipsum}
\usepackage[normalem]{ulem}
\usepackage[percent]{overpic}
\usepackage{color}

\begin{document}
\title{Inverse engineering of cooling protocols: from normal behavior to Mpemba effects}

\author{Hartmut L\"owen}
\affiliation{Institut f\"ur Theoretische Physik II: Weiche Materie, Heinrich-Heine-Universit\"at D\"usseldorf, Universit\"atsstra{\ss}e 1, 
40225 D\"usseldorf, Germany}

\begin{abstract}
When a cup of hot coffee is suddenly put into a cold environment, it cools down as a function of time $t$ until the internal temperature $T_\text{int}$ of the coffee equals the external ambient temperature $T_\text{ext}$. This instantaneous shock-freezing corresponds to an imposed cooling protocol of the external temperature $T_\text{ext}(t)$, ideally described as a step-function in time,  causing the time-dependent change of the internal temperature $T_\text{int}(t)$. While the effect of different given protocols $T_\text{ext}(t)$ on the resulting system cooling behaviour, embodied in $T_\text{int}(t)$, has been studied extensively, we consider here the inverse question: for a given system cooling $T_\text{int}(t)$ how can an appropriate protocol $T_\text{ext}(t)$ be engineered to produce the desired prescribed $T_\text{int}(t)$. We use both the phenomenological Newtonian equation for cooling and microscopic models, such as a discrete two-level system and a Brownian harmonic oscillator with time-dependent noise, to compute analytically the protocol $T_\text{ext}(t)$ needed to achieve a prescribed $T_\text{int}(t)$. We then discuss the same question for phenomenological generalizations of the Newtonian law which include anomalous Mpemba effects, overcooling, asymmetries in cooling and heating as well as delay phenomena. It is shown that backward-engineered protocols do not always exist and can be non-unique. The results are important for steering the cooling behavior by time-varying external heat sources in a systematic way.
\end{abstract} 

\maketitle

\section{introduction}
Cooling and heating processes are known from everyday life ranging from cooking of eggs \cite{egg_cooking} over melting of cheese \cite{Mathijssen_RMP} to the transmission of heat through the human skin \cite{skin} but they are also relevant for many applications in industry such as steel production \cite{review_steel}, glass formation \cite{glasses} and emerging energy technology \cite{review_combustion}. A fundamental scientific treatment of cooling involves nontrivial concepts from non-equilibrium statistical physics. One of the simplest phenomenological description is the traditional Newtonian cooling law \cite{Scala_1701,Newton2,Newton3}: If a system with an internal temperature $T_\text{int}$ is brought in contact with an external bath at fixed temperature $T_\text{ext}$, the cooling rate of the system is proportional to the actual temperature difference $T_\text{ext}-T_\text{int}$. This will drive the system temperature exponentially in time $t$ towards the ambient temperature until there is no further heat exchange between the system and the bath.

Real cooling phenomena, however, can be more complicated than Newton’s law \cite{Newton3,limit2}. One important example is the Mpemba effect, originally discovered for water, which refers to the counter-intuitive finding that hot systems can cool faster than warm ones. The Mpemba effect was first reported in antiquity in Aristotle’s Meteorologica \cite{Aristoteles_1952} who stated “to cool hot water quickly, begin by putting it in the sun”. In 1969 Mpemba and Osborne observed this unexpected cooling behavior while freezing ice cream in a refrigerator \cite{Mpemba_1969}. In modern nomenclature, the Mpemba effect refers to a phenomenon in which a system initially at a higher temperature $T_\text{int}$ cools faster than the same system started from a lower temperature when both are quenched into the same cold environment at $T_\text{ext}$. This cannot be explained by the classical Newtonian cooling law. In more general thermodynamic terms, the existence of a Mpemba effect does imply that the system deviates from a quasi-static cooling path close to equilibrium, it rather retains true memory about its cooling history. 
Controversial and counter-intuitive by nature, the Mpemba effect has attracted broad attention since it has been observed in a wide range of systems. Beyond the canonical case of water freezing where the existence of the Mpemba effect is still controversial (see \cite{Teza_2025} for a recent review) notable instances include but are not limited to colloidal suspensions \cite{Kumar_2020,Kumar_2022,Malhotra_2024}, granules \cite{Lasanta_2017,Biswas_2023,Megias_2022} and quantum spin systems \cite{Chatterjee_2023,Nava_2024,Joshi_2024,Mororder_2024,Xu_2025}. Nevertheless, despite its consistent phenomenological signature, the underlying microscopic mechanisms appear to be system-specific and remain the subject of an intense ongoing debate \cite{Bechhoefer_2021}.

Here we shall generalize cooling phenomena to general time-dependent external heat bath temperature protocols $T_\text{ext}(t)$ exerted from the bath onto the system. Optimizing egg-cooking where the external temperature is not constant nor shock-heated but actually time-dependent \cite{egg_cooking} is one prominent example. The standard protocol typically studied in the context of the Mpemba effect is an instantaneous temperature shock at time $t=0$ where the temperature of the bath quickly jumps from a temperature equilibrated with the system, $T_\text{ext}(t=0) = T_\text{int}(t=0)$ towards a smaller target temperature $T_\text{ext}(\infty )$. In this case, the external temperature protocol $T_\text{ext}(t)$ can be idealized by a step function, i.e. 
\begin{equation}
    \label{eq:1}
    T_\text{ext}(t) = T_\text{ext}(0) + \Theta(t)(T_\text{ext}(\infty ) – T_\text{ext}(0) ).
\end{equation}

Research has typically addressed the impact of different time-dependent protocols $T_\text{ext}(t)$ on the system cooling curve $T_\text{int}(t)$. Or, in other terms, the question is how a time-dependent protocol influences the cooling behaviour of the system \cite{Teza_2025}. Key examples are a constant cooling rate imposed externally on the system \cite{Santos_2025} as relevant for glass formation \cite{cooling_glasses} or a multi-step cooling/heating protocol \cite{Raz_2020} required for the Kovacs \cite{Kovacs1,Kovacs2,Kovacs3,Kovacs4} and the Pontus-Mpemba effect \cite{Nava_2025}. 

In this paper, we reverse this question. Instead of computing the cooling curve of the system, we prescribe it as a given input and  compute the protocol needed to achieve this prescribed cooling curve. Obviously this backward question is central if the full system temperature development wants to be controlled and steered as a function of time. In this way, the protocol $T_\text{ext}(t)$ needed to produce the desired prescribed cooling curve $T_\text{int}(t)$ is determined by {\it inverting} the relation between $T_\text{ext}(t)$ and $T_\text{int}(t)$. We perform this inversion explicitly for the phenomenological Newtonian cooling as well as for two microscopic models, namely a two-level system with detailed balance hopping rates \cite{detailed_balance} and a Brownian oscillator whose noise strength is adjusted to a prescribed system temperature \cite{Zerbe_1994,Brouard_2001,Barzykin_1998,Berdichevsky_1999,Gammaitoni_1998,Schmiedel_2008}. It should be emphasized that such inverse engineering of protocols has been considered in other contexts earlier, e.g.\ for quantum short-cuts to adiabaticity \cite{Muga}, for a Brownian particle in a controllable harmonic potential  \cite{Ciliberto,Pires}, for relaxation optimization \cite{Prados_2021,Patron} and for active particles \cite{Davis,Olsen2}, see \cite{review_Trizac} for a review.

Then we introduce appropriate simple generalizations of the Newtonian description which include Mpemba effects, overcooling, asymmetries in cooling and melting, as well as delay phenomena. Within the framework of these generalized simplistic models, engineered protocols are determined for prescribed system cooling curves. We then turn to existence and uniqueness of the inverse-engineered protocol. 
In fact, backward-engineered protocols do not always exist, in particular if drastic system cooling rates should be achieved \cite{Pires}. For nonlinear models, e.g.\ valid for materials with a negative differential heat conductivity, the inverse-engineered protocols are not unique such that there are even multiple solutions for the protocol.  Our results not only give insight into the underlying physics of cooling but are also important for steering the cooling behavior by time-varying external heat baths in a systematic way.

The paper is organized as follows: in section II, we first study normal cooling behaviour within the traditional Newtonian approach. We then consider two simple microscopic models where an analytical solution of the inversion is possible. These are a simple discrete two-level system where the transition rates are changing externally as a function of time and a Brownian harmonic oscillator where the noise strength is varied as a function of time.  Then, in section III, we turn to phenomenological models which include anomalous cooling as described by a suitable generalization of the Newtonian cooling law. Then we explore the existence and uniqueness of the backward engineered protocol in sections IV and V. Conclusions are presented finally in section VI.


\section{Engineered protocols for normal cooling}
We first illustrate the inverse engineering concept for normal cooling which does not exhibit any Mpemba effect. The details depend of course on how the coupling between the system (internal) and the environment (external) is realized and how the non-equilibrium system temperature is defined precisely. For the simplest case of Newtonian cooling it is just the bulk temperature difference that is the phenomenological driving force. For the two level model, the system temperature is defined by matching the occupation number to the associated thermal equilibrium Boltzmann distribution while the coupling to the external bath is realized via the transition rates between the two levels which fulfill the detailed balance condition imposed by the external temperature. Third, for the Brownian oscillator, the coupling is defined via the noise strength of the solvent which performs thermal kick on the particle and the system temperature is matched by mapping its density distribution to an actual thermal Gaussian one. 

\subsection{Newtonian cooling}
\begin{figure}[htbp]
    \centering
    \includegraphics[width=0.99\linewidth]{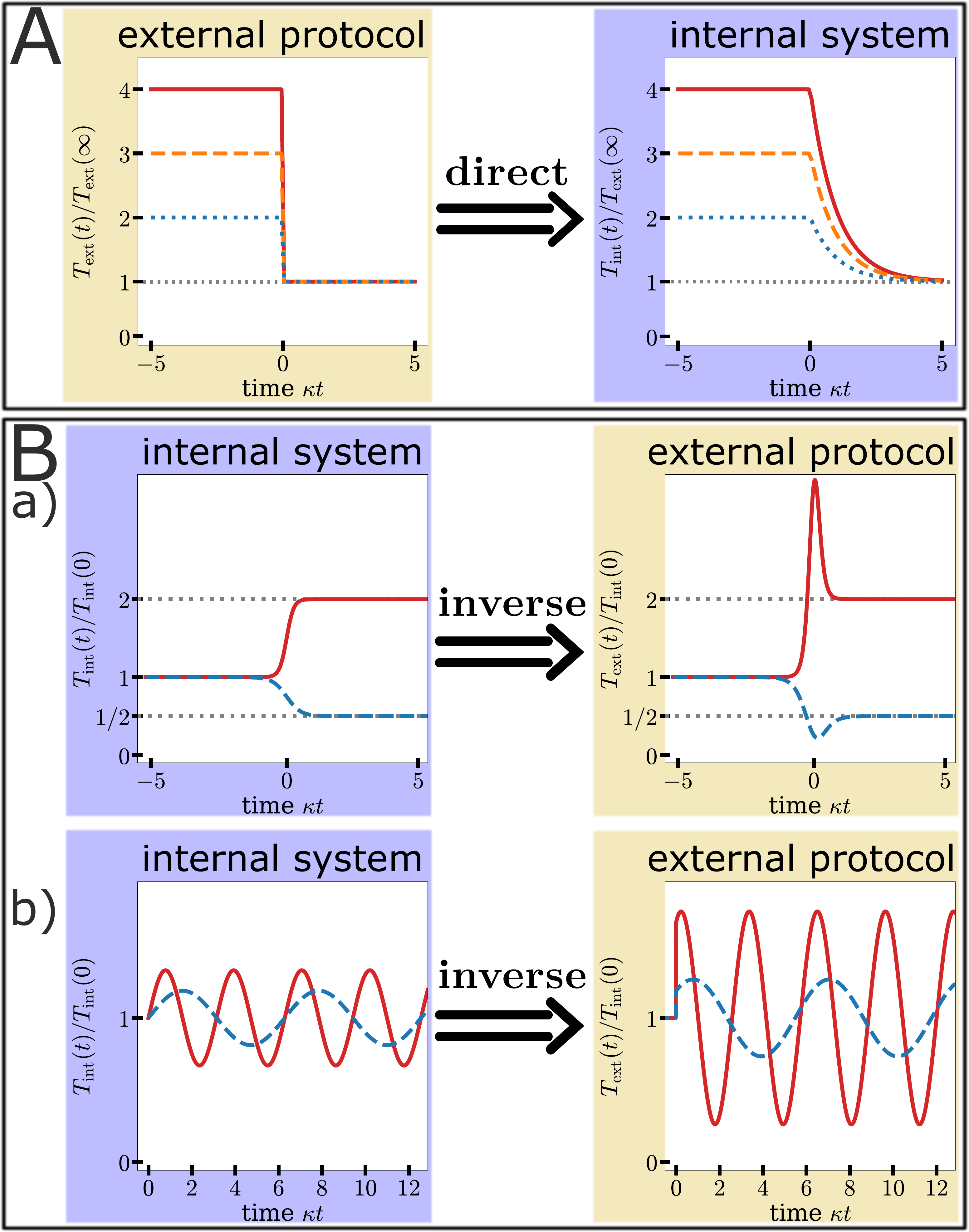}
    \caption{Correspondence between the external bath temperature protocol $T_\text{ext}(t)$ and the internal system temperature $T_\text{int}(t)$ for normal cooling based on the Newtonian law which involves a cooling rate $\kappa$. 
    A)	{\bf Direct} way: application of the protocol $T_\text{ext}(t)$ (left panel) produces the system cooling shown as $T_\text{int}(t)$ (right panel). Three different protocols which a cooling shock at $t=0$  are shown for three different initial temperatures $T_\text{ext}(0)$. 
    B)	{\bf Inverse} way: different desired system cooling curves $T_\text{int}(t)$ (now in the left panel) and the protocol $T_\text{ext}(t)$ (right panel) needed to generate them. a) Shock-heating and more gentle cooling at a time $t_0$ starting from thermal equilibrium for $t=0$.  b) Two cyclic cooling and heating  curves $T_\text{int}(t)$ (left panel) and their required oscillatory protocol $T_\text{ext}(t)$ (right panel).}
    \label{fig:1}
\end{figure}

The traditional Newtonian cooling equation reads as
\begin{equation}
    \label{eq:2}
    {\dot T}_\text{int} (t) = -\kappa ( T_\text{int} (t) – T_\text{ext} )
\end{equation}
where the dot refers to a derivative in time $t$ and $T_\text{ext}$ is the constant target temperature the system is exposed to. Moreover $\kappa$ represents an effective system cooling rate when coupled to an external environment. This parameter $\kappa$ is assumed to be constant and sets the inverse time scale. $\kappa$ can depend on many details, in particular on the heat conductivity realized in the coupling between the external and internal system. The solution of \eqref{eq:2} for $T_\text{int}(t)$ with an initial temperature $T_\text{int}(0)$ at time $t=0$ is simply decaying exponentially in time with the rate $\kappa$
\begin{equation}
    \label{eq:3}
    T_\text{int}(t) = (T_\text{int}(0) –T_\text{ext}) \exp (-\kappa t) + T_\text{ext}.
\end{equation}
Invoking the adiabatic approximation \cite{Santos_2025}, one can readily extend Eq.\ \eqref{eq:2} for time-dependent protocols $T_\text{ext}(t)$ towards the equation
\begin{equation}
    \label{eq:4}
    {\dot T}_\text{int} (t)  = -\kappa ( T_\text{int} (t) – T_\text{ext}(t) )
\end{equation}
which has the solution
\begin{equation}
    \label{eq:5}
    T_\text{int}(t) = T_\text{int}(0) \exp (-\kappa t) + \int_0^t  dt’ \kappa T_\text{ext}(t’) \exp ( - \kappa (t-t’)).
\end{equation}
This direct way is illustrated in Figure~\ref{fig:1}A. An application of a protocol $T_\text{ext}(t)$ (left panel) with three different initial temperatures $T_\text{ext}(0)$ produces the system cooling curves shown as $T_\text{int}(t)$ (right panel). For $t\leq 0$ the system and the environment are in thermal equilibrium, $T_\text{int}(0) = T_\text{ext}(0)$. Clearly, the resulting system temperature curves $T_\text{int}(t)$ are monotonic in the initial temperature $T_\text{int}(0)$ which implies that there is no Mpemba effect.

Next we perform the inversion between $T_\text{int}(t)$ and $ T_\text{ext}(t)$ by expressing $T_\text{ext}(t)$ for a given $T_\text{int}(t)$. Using Eq. \eqref{eq:4}, which can directly be solved for $T_\text{ext}(t)$ the backward solution is simply given by
\begin{equation}
    \label{eq:6}
     T_\text{ext} (t)  = T_\text{int} (t) +  {\dot T}_\text{int} (t) / \kappa.
\end{equation}
Mathematically $T_\text{ext} (t)$  is expressed by \eqref{eq:6} as a {\it functional} of $T_\text{int} (t)$. 

In order to illustrate the results of the inversion, let us insert some examples. If one looks for a protocol that produces a shock-freeze of the system from an initial temperature $T_\text{int}(0)$ to a final temperature $T_\text{int}(\infty )$ at time $t=0$ - which should be contrasted to the shock-freeze of the environment as written in Eq.\ \eqref{eq:1} -  the desired internal system temperature is
\begin{equation}
    \label{eq:7}
    T_\text{int} (t) = T_\text{int} (0) + \theta ( t )  (T_\text{int} (\infty) – T_\text{int}(0) ).
\end{equation}
By inserting this into \eqref{eq:6} the unique engineered protocol $T_\text{ext} (t)$  which produces this system shock-frozen cooling behaviour is
\begin{equation}
    \label{eq:8}
    \begin{aligned}
    T_\text{ext} (t) &=      T_\text{int} (0)  + \theta ( t)  (T_\text{int} (\infty) – T_\text{int}(0) ) \\
    &+ \delta ( t )  (T_\text{int} (\infty) – T_\text{int}(0) )  / \kappa
    \end{aligned}
\end{equation}
such that a desired shock-frozen behavior is generated by a $\delta$-spike in the protocol temperature. This reveals a fundamental difference between cooling and heating: while for heating any high external temperature does exist to realize the positive $\delta$-spike, this is not the case for the negative $\delta$-spike required for cooling since the external temperature cannot become negative.  We shall discuss this in more detail in section IV.

In Figure~\ref{fig:1}B, Eq.\ \eqref{eq:6} is illustrated for two situations, namely, the engineered protocols required to produce:  a) shock heating (red curve) and gentle cooling (blue curve), and b) two oscillatory system temperatures sinusoidal in time, as considered recently in Ref.\ \cite{Abreu}. Quick heating needs a strong increase in the system temperature which exceeds the final target temperature. Gentle cooling of the system requires also a temperature undershoot but strong cooling rates are prohibited since the external temperature cannot become negative. The two cyclic cooling and heating curves $T_\text{int}(t)$ shown in Figure~\ref{fig:1}B b) (left panel)  require a phase-shifted oscillatory protocol $T_\text{ext}(t)$ with a higher amplitude (right panel).

\subsection{Two-level system}
As one of the simplest microscopic toy model, we introduce a discrete two-level system. Such a model is frequently used for the thermalization of a quantum $1/2$ spin. Without loss of generality let us assume that the ground state $A$ has zero energy while the second excited state $B$ possesses an eigen-energy $\epsilon > 0$. The time-dependent system temperature $T_\text{int} (t)$ is linked to the instantaneous associated Boltzmann distribution. In a two-level system, it is sufficient to match just the single probability $P_A(t)$ to occupy the ground state $A$ since the occupation of the second state is then fixed by the complementary probability. This yields
\begin{equation}
    \label{eq:2b_1}
    P_A(t) = \frac {1}{1+\exp (- \epsilon/k_BT_\text{int}(t))}
\end{equation}
where $k_B$ denotes Boltzmann’s constant. Conservation of the total probability leads to the constraint 
\begin{equation}
    \label{eq:2b_2}
    P_B(t) = 1-P_A(t)
\end{equation}
for the occupation probability $P_B(t)$ of the excited state $B$ at all times $t$.
The cooling dynamics is described by the Master equation
\begin{equation}
    \label{eq:2b_3}
    {\dot P_A(t)} = - \Gamma_{A\to B} P_A(t) + \Gamma_{B\to A}(t) P_B(t) .
\end{equation}
Here the time-independent transition rate $\Gamma_{A\to B}$ sets the time scale and the time-dependent rate $\Gamma_{B\to A}(t)$  defines via the detailed balance condition 
\begin{equation}
    \label{eq:2b_4}
    \Gamma_{B\to A}(t) = \Gamma_{A\to B} \exp (\epsilon/ k_B T_\text{ext}(t))
\end{equation}
at any time the external temperature $T_\text{ext}(t)$. Or in other words, the transition rates are determined by the external bath temperature while the actual occupation ratio sets the internal system temperature $T_\text{int}(t)$. In case the external temperature is constant, the Master equation \eqref{eq:2b_3} converges to the Boltzmann distribution for long times such that external and internal temperature are constant and coincide which is the condition of thermodynamic equilibrium. However for time-dependent protocols, these two temperatures differ in general.

\begin{figure}[htbp]
    \centering
    \includegraphics[width=0.99\linewidth]{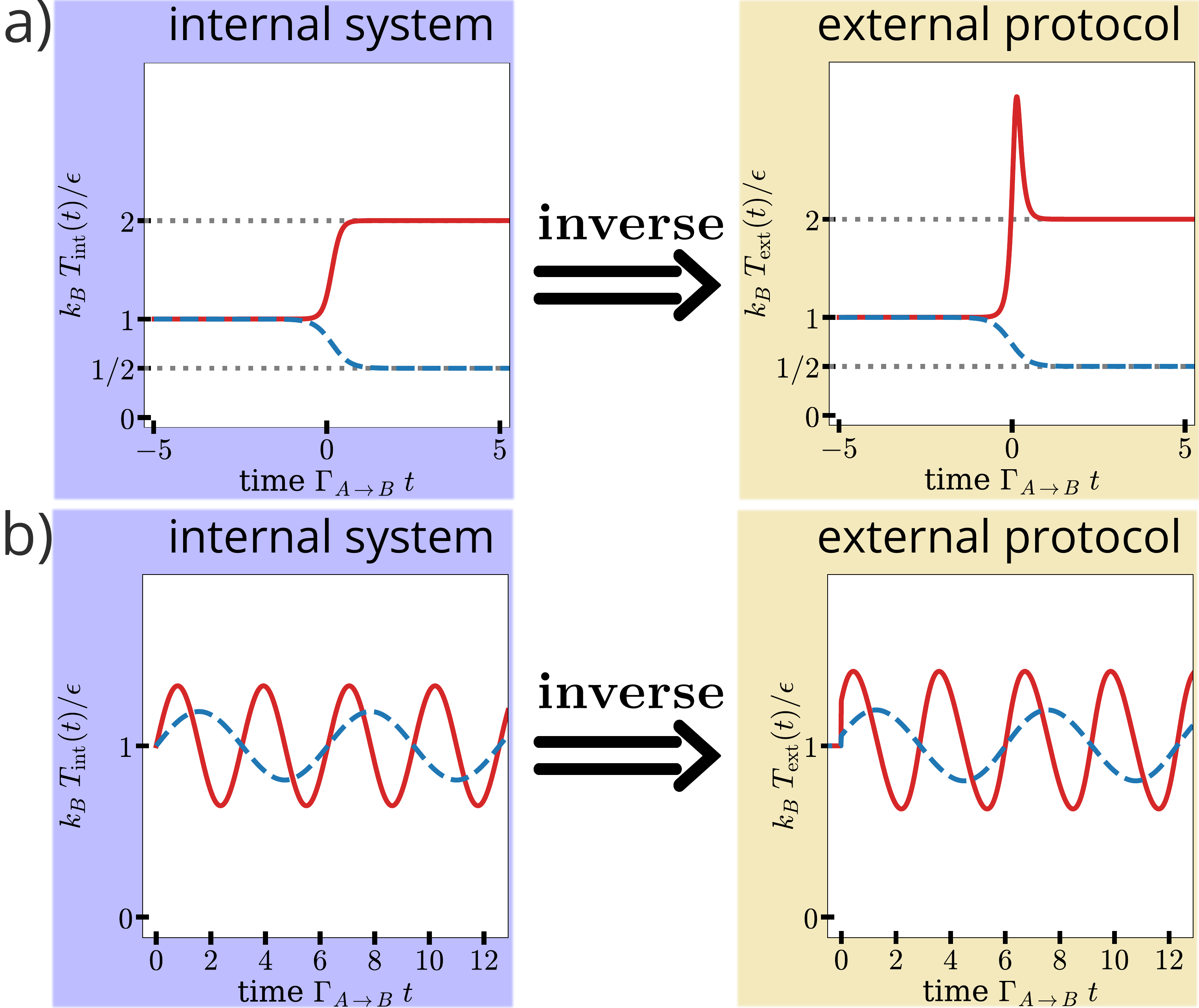}
    \caption{{\bf Inverse} way: Same as Figure~\ref{fig:1}B but now for the two-level system. Cooling curves of the system (left panel) and the protocols needed to achieve those (right panel). The time unit is $1/\Gamma_{A\to B} $ and the temperature unit is $\epsilon/k_B$.
	a) Shock-like heating and more gentle cooling curves $T_\text{int}(t)$ of the system and the associated protocol $T_\text{ext}(t)$. b) Cyclic cooling curves $T_\text{int}(t)$ corresponding to an oscillatory protocol $T_\text{ext}(t)$.}
    \label{fig:2}
\end{figure}

Combining Eqns. \eqref{eq:2b_3} and \eqref{eq:2b_2} we obtain the inversion for the engineered protocol and express $T_\text{ext}(t)$ in terms of $T_\text{int}(t)$ as follows
\begin{equation}
    \label{eq:2b_5}
    T_\text{ext}(t) = \frac {\epsilon}{k_B}    \frac {1}{ \ln \left[ \frac{1}{E(t)} + \frac {\epsilon \ \dot T_\text{int}(t)}{ \Gamma_{A \to B}(1+E(t))k_B T^2_\text{int}(t) } \right] }
\end{equation}
with the Boltzmann factor $E(t)= \exp (-\epsilon/k_BT_\text{int}(t) )$. We have evaluated Eqn.\ \eqref{eq:2b_5} in Figure~\ref{fig:2} for the same different desired system temperatures as those used in Figure~\ref{fig:1} and the results are qualitatively similar. In particular there is no Mpemba effect. For small  gradients 
in $T_\text{int}(t)$, Eqn.\ \eqref{eq:2b_5} can be linearized and the same qualitative equation is obtained as for the phenomenological Newtonian cooling law. However, when comparing Figure~\ref{fig:2} with Figure~\ref{fig:1} there is no overshooting in the gentle cooling law for the two-level system cooling and the temperature spike needed for drastic heating is smaller in amplitude.

\subsection{Brownian harmonic oscillator}
Harmonic traps, as realized by optical tweezers for Brownian colloidal particles \cite{Volpe_2023,Buttinoni_2022} and granules \cite{Demery}, are common confinements and ideal to study non-equilibrium dynamics of small systems.  Here we consider a single Brownian particle moving in one spatial dimension with its trajectory $x(t)$. The Brownian harmonic oscillator  \cite{Loewen_JPCM,Seifert_2011} is then defined by the following stochastic overdamped equation of motion
\begin{equation}
    \label{eq:2c_1}
    \gamma {\dot x(t)} = - \lambda x(t) + f(t)
\end{equation}
where $\gamma$ is a friction coefficient and $\lambda$ denotes the trap strength or spring constant. The stochastic forces $f(t)$ are Markovian Gaussian random numbers of zero mean, ${\overline {f(t)}}=0$, and variance 
\begin{equation}
    \label{eq:2c_2}
    {\overline {   f(t)  f(t’)  }       } = 2\gamma k_B T_0  \delta ( t-t’)
\end{equation}
with the overbar denoting a stochastic average over the noise. Here $T_0$ is the external bath temperature provided by the thermal noise of the embedding solvent. The fluctuation-dissipation relation \eqref{eq:2c_2} is strictly speaking only valid in equilibrium when the external temperature $T_0$ of the solvent is constant. 

This Brownian harmonic oscillator has already been used as a "working horse" to discuss inverse-engineered protocols  both for a time-dependent stiffness $\lambda (t)$ \cite{Ciliberto,review_Trizac,NPJ_Trizac,Ibanez}, for a time-dependent solvent temperature \cite{Pires} or for both \cite{Chupeau_2}. Let us here recapitulate the inverse-engineered protocol if only the solvent temperature is varied.

The linear stochastic equations \eqref{eq:2c_1} of motion can be solved analytically, see e.g. \cite{Loewen_JPCM}. The equilibrium distribution of the particle is Gaussian with a variance 
\begin{equation}
    \label{eq:2c_3}
    \Delta^2:={\overline {x^2(t)}} =  k_B T_0 / \lambda.
\end{equation}
This width turns out to be an ideal parameter to characterize the internal system temperature. 

Now we generalize the equation of motion \eqref{eq:2c_1} to a time-dependent external temperature $T_\text{ext}(t)$. Accordingly we now model the stochastic forces with a time-dependent second moment by
\begin{equation}
    \label{eq:2c_4}
    {\overline {   f(t)  f(t’)  }       } = 2\gamma k_B T_\text{ext}(t)  \delta ( t-t’).
\end{equation}
Then the equations of motion can still be solved resulting in a Gaussian density distribution with a time-dependent variance 
\begin{equation}
    \label{eq:2c_5}
    \Delta (t)^2 = \Delta (0)^2 + 2 \exp (-2\omega_0 t) k_B \int_0^t dt’ T_\text{ext}(t’) \exp (2\omega_0 t’)/\gamma
\end{equation}
where $\omega_0 = \lambda/\gamma$ denotes the damping eigenfrequency of the Brownian harmonic oscillator. Generalizing the relation \eqref{eq:2c_3}, we define a time-dependent internal temperature $T_\text{int}(t)$ by matching the Gaussian density distribution to an appropriate equilibrium one via
\begin{equation}
    \label{eq:2c_6}
    T_\text{int}(t) = \lambda \Delta(t)^2 /k_B
\end{equation}
such that
\begin{equation}
    \label{eq:2c_7}
T_\text{int}(t) = 2 \omega_0 \exp (-2\omega_0 t)  \int_0^t dt’ T_\text{ext}(t’) \exp (2\omega_0 t’).
\end{equation}
Inverting this relation yields 
\begin{equation}
    \label{eq:2c_8}
    T_\text{ext}(t) =T_\text{int}(t) + {\dot T}_\text{int}(t)/2\omega_0.
\end{equation}
Interestingly, this has the same algebraic form as for the Newtonian cooling in Eq.\ \eqref{eq:6} where the cooling rate coefficient $\kappa$ is played by the doubled eigenfrequency of the oscillator, hence the same conclusions apply.  For the harmonic Brownian oscillator there is no Mpemba effect and the engineering of cooling curves can be read off from Figure~\ref{fig:1}. However, here we have obtained the Newtonian cooling law for a microscopic model. It should be remarked the Newtonian cooling law has been derived also based on much detailed models using linear response theory \cite{derive_Newton1,limit2,Chinese_2026}. 


\section{Engineered protocols for anomalous cooling}
The Newtonian cooling law can be extended into several directions. Here we propose simple phenomenological models in order to constitute a framework to accommodate a variety of anomalous cooling effects including Mpemba effects, overcooling, asymmetry between heating and cooling and time-delay. The traditional Newtonian cooling is always retained as a special limit.

\subsection{Mpemba effects}
First we shall explore the dependence of the cooling rate $\kappa$ on the initial internal temperature $T_\text{int}(0)$. This simple phenomenological dependence can be used to model strong Mpemba effects. For a cooling rate $\kappa (T_\text{int})$, the solution presented in Eq.~\eqref{eq:2} for a constant target temperature $T_\text{ext}$ generalizes to
\begin{equation}
    \label{eq:3a_1}
    {\dot T}_\text{int} (t) = -\kappa (T_\text{int}(0)) ( T_\text{int} (t) – T_\text{ext} ).
\end{equation}
If the rate $\kappa  (T_\text{int}(0)) $ increases with temperature $T_\text{int}(0)$, a system at an initial higher temperature will decay faster towards the target temperature than an initial warm one which is the strong Mpemba effect. 
By non-monotonicities in $\kappa  (T_\text{int}(0))$ even more subtle behavior, like multiple Mpemba effects \cite{Malhotra_2024} can be modelled and almost the whole zoo of Mpemba effects \cite{Scala_1701} can phenomenological incorporated in an appropriate choice of $\kappa  (T_\text{int}(0))$.

\begin{figure}[htbp]
    \centering
    \includegraphics[width=0.55\linewidth]{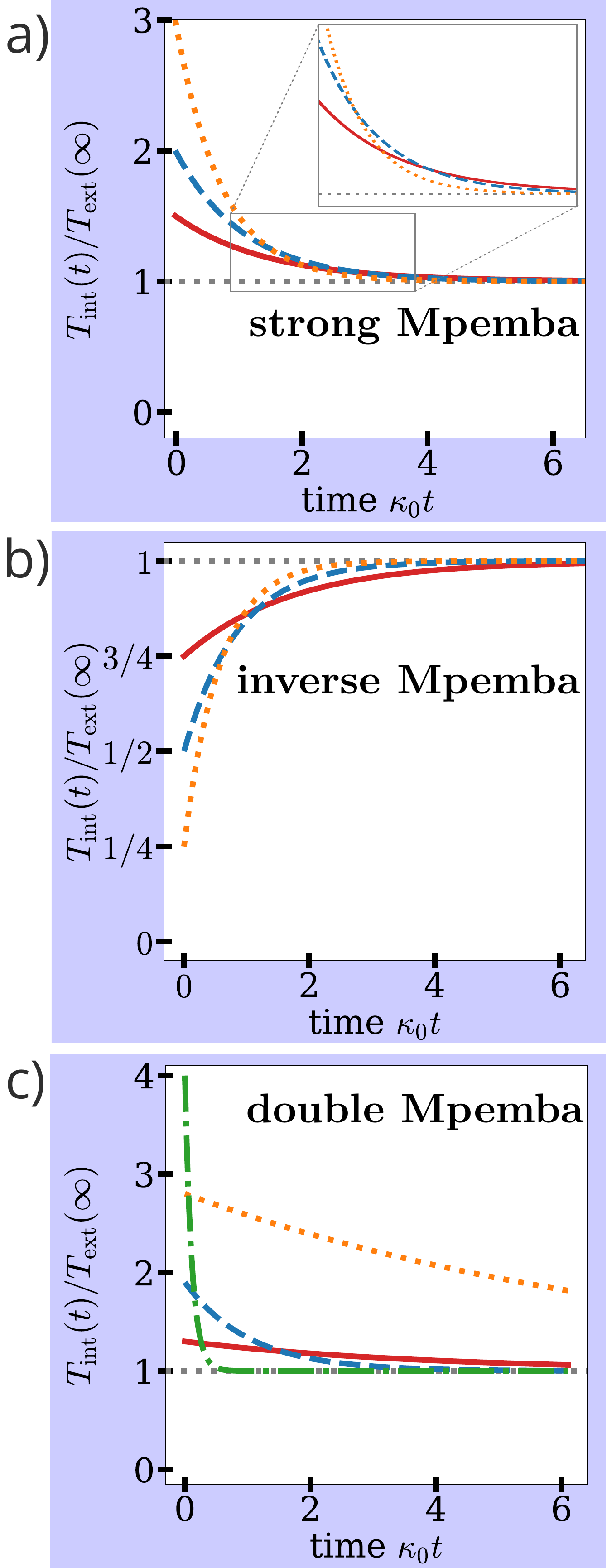}
    \caption{System cooling behavior $T_\text{int}(t)$ upon shockwise freezing from an initial temperature $T_\text{int}(0)$ to an external  target temperature $T_\text{ext}(\infty )$ at $t=0$. Units of temperature and time are  $T_\text{ext}(\infty )$ and $1/\kappa_0$   a) {\bf Strong Mpemba effect} for  $\kappa  (T_\text{int}(0))$ as given in Eq.\ \eqref{eq:3a_1x} with $\alpha_1 = 1/T_\text{ext}(\infty )$. The inset magnifies the relevant time domain where the higher initial temperature overtakes cooling relative to a warm temperature. b) {\bf Inverse Mpemba effect} for  $\kappa  (T_\text{int}(0))$ as given in Eq.\ \eqref{eq:3a_1y} with $\alpha_2=1/T_\text{ext}(\infty )$. c) {\bf Double Mpemba effect} for $\kappa  (T_\text{int}(0))$  as given in Eq.\ \eqref{eq:3a_1z} with $T_c=2 T_\text{ext}(\infty )$ and $T_0= T_\text{ext}(\infty ) $.}
    \label{fig:3}
\end{figure}

Three examples for the system cooling curve upon a sudden quench are shown in Figure~\ref{fig:3} for the following functions: 
\begin{enumerate}
    \item {\bf Strong} Mpemba effect for a linear dependence 
    \begin{equation}
        \label{eq:3a_1x}
        \kappa  (T_\text{int}(0))  = \kappa_0 (1 + \alpha_1 T_\text{int}(0))
    \end{equation}
    with both positive $\kappa_0$ and $\alpha_1$.
    \item {\bf Inverse} Mpemba effect for the linear dependence 
    \begin{equation}
        \label{eq:3a_1y}
        \kappa  (T_\text{int}(0))  = \kappa_0 (1 - \alpha_2 T_\text{int}(0))
    \end{equation}
    of $\kappa  (T_\text{int}(0))$ with both positive $\kappa_0$ and $\alpha_2$. This is for heating, hence a target temperature $T_\text{ext}$ higher than the initial one is imposed.
    \item {\bf Double} Mpemba effect for a phenomenological choice of
    \begin{equation}
        \label{eq:3a_1z}
        \kappa  (T_\text{int}(0))  = \kappa_0  ( (T_\text{int}(0)-T_c)^2  - T_0^2)^2/T_0^4 
    \end{equation}
    with positive $\kappa_0$, $T_c$ and $T_0$. Due to the non-monotonic variation of $\kappa  (T_\text{int}(0))$ (resp. the associated cooling time) multiple Mpemba effects can be induced.
    \end{enumerate}

\begin{figure}[htbp]
    \centering
    \includegraphics[width=0.99\linewidth]{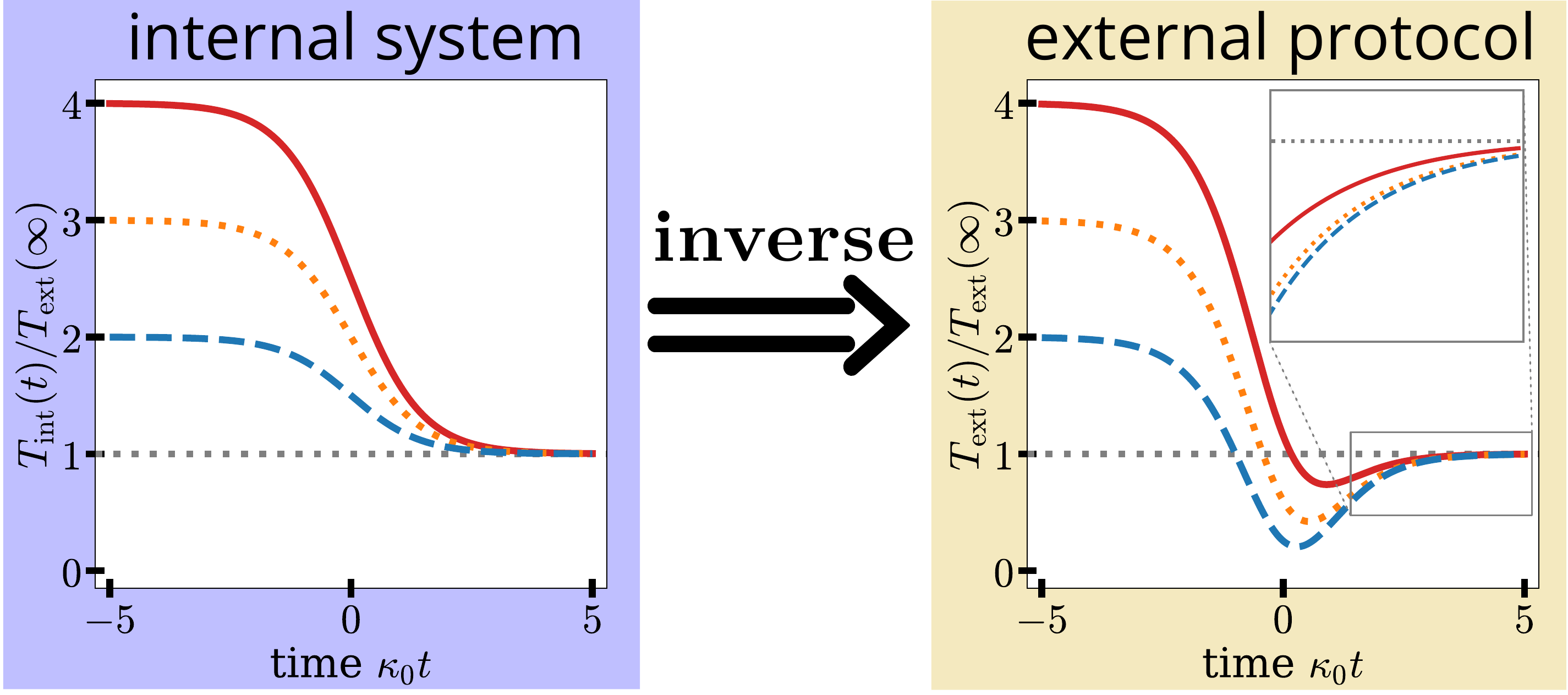}
    \caption{Same as Figure~\ref{fig:1}B a) but now for the strong Mpemba effect where units of temperature and time are  $T_\text{ext}(\infty )$ and $1/\kappa_0$   and  $\kappa  (T_\text{int}(0))  = \kappa_0(1 + \alpha_1 T_\text{int}(0))$ with $\alpha_1=1/T_\text{ext}(\infty )$. The inset magnifies the relevant time domain where the higher initial temperature achieves quicker cooling relative to a warm temperature.}
    \label{fig:4}
\end{figure}

The inverted equation now reads as 
\begin{equation}
    \label{eq:3a_2}
    T_\text{ext}(t) = T_\text{int} (t) +    {\dot T_\text{int} (t) } / \kappa (T_\text{int} (0).
\end{equation}
Some engineered cooling protocols are shown in Figure~\ref{fig:4} demonstrating that now the details of these protocols depends on the initial temperature of the system. For a similar desired system cooling rate, the higher the initial temperature the smaller the undershoot needed which follows from the fact that the magnitude of the undershoot scales with $1/\kappa  (T_\text{int}(0))$. In parallel, the strong Mpemba effect ensures that the time needed to perform the undershoot in the engineered protocol is shorter for higher initial temperatures, see the inset in Figure~\ref{fig:4}.

\subsection{Overcooling}
Transient overcooling where the transient temperature is lower than the imposed target temperature has recently been found in several systems including spins \cite{Parisi_PNAS}, confined active matter \cite{Schwarzendahl_2022} and frictional granules \cite{Antonov_2026}. Overcooling can be included when taking additionally inertial memory into account 
\begin{equation}
    \label{eq:3b_1}
    \mu {\ddot T_\text{int} (t) } + {\dot T_\text{int} (t) } = -\kappa  ( T_\text{int} (t) – T_\text{ext}(t) )
\end{equation}
where $\mu$ is a (formal) mass. The inertia here is phenomenological and should be distinguished from inertia in the microscopic equation of motion governing individual particle trajectories. The equation \eqref{eq:3b_1}  possesses the form of a damped harmonic oscillator. It can be microscopically derived for an overdamped particle in a harmonic potential with run-and-tumble dynamics (or socalled telegraphic noise) \cite{Dhar}. For small inertia we are in the overdamped regime and for a shockwise freezing towards a constant target temperature $T_\text{ext}$ the solution is 
\begin{equation}
    \label{eq:3b_2}
    T_\text{int} (t) = {\bar A} \exp (- \gamma_+ t)   + {\bar B}   \exp (-\gamma_- t) + T_\text{ext}
\end{equation}
with $\gamma_{\pm}=-\beta \pm \sqrt{ \beta^2 - \Omega_0^2}$, $\beta = 1/{2\mu}$, $\Omega = \sqrt{\kappa/\mu}$, $ {\bar A}=  T_\text{int}(0) – T_\text{ext} –{\bar B}  $
and ${\bar B}=((T_\text{ext} – T_\text{int}(0)\gamma_+ – {\dot T}_\text{int}(0))/(\gamma_+ + \gamma_-)$.
Here the initial cooling rate ${\dot T}_\text{int}(0)$ of the system enters explicitly as an additional parameter which introduces some kind of memory \cite{NPJ_Trizac}.  

\begin{figure}[htbp]
    \centering
    \includegraphics[width=0.99\linewidth]{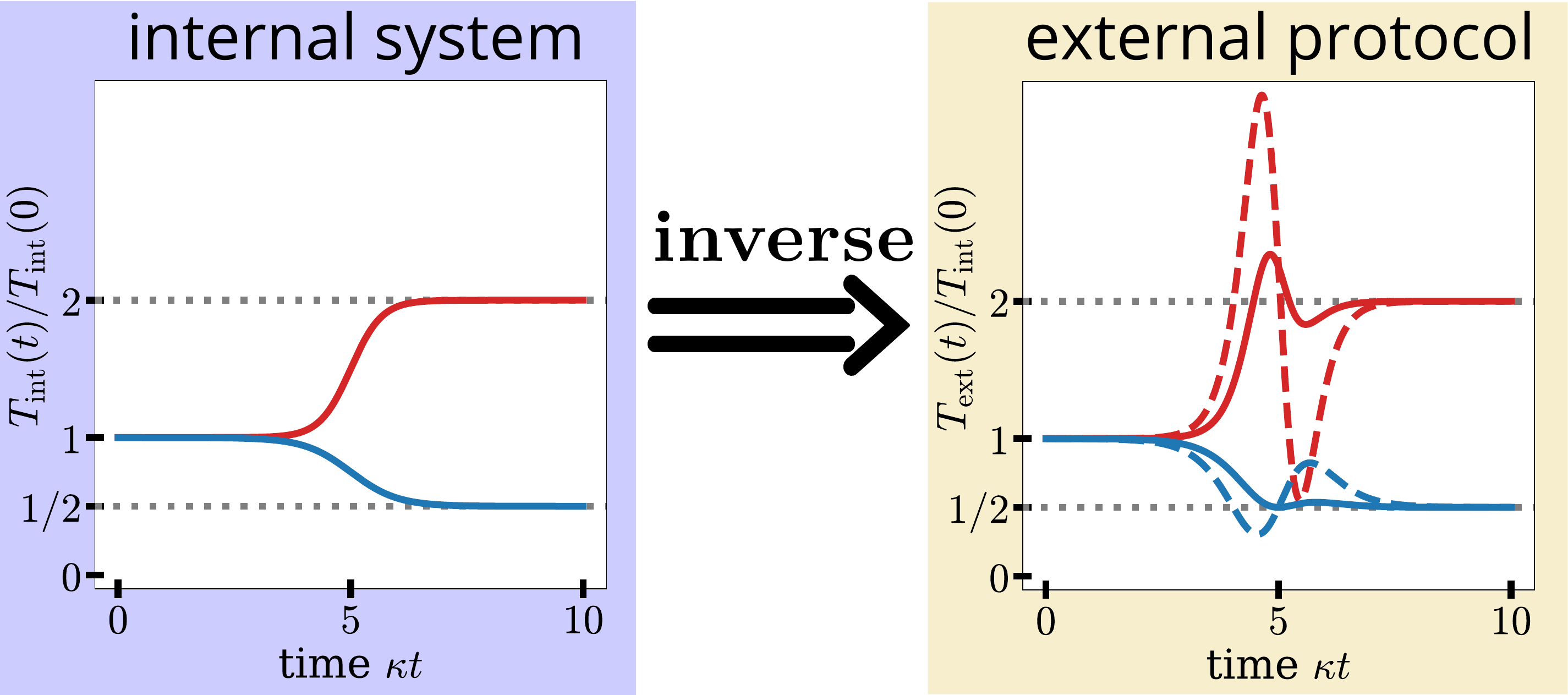}
    \caption{Same as Figure~\ref{fig:1}B a) but now for the inertial cooling equation. The time unit is $1/\kappa$ and the temperature unit is $T_\text{int}(0)$ with $\mu=0.5$
    (solid line) and $\mu=2$
    (dashed line).}
    \label{fig:5}
\end{figure}

The inverted relation between $T_\text{int} (t)$ and  $T_\text{ext}(t)$ is given by
\begin{equation}
    \label{eq:3b_3}
    T_\text{ext}(t) = T_\text{int} (t) +   ({\dot T_\text{int} (t) } + \mu {\ddot T_\text{int} (t) }) / \kappa.
\end{equation}
Some examples for inverse-engineered protocols for cooling and heating are shown in Figure~\ref{fig:5}. They exhibit a significant oscillatory behavior in the cooling protocols which depend on the amount of inertia. 

\subsection{Asymmetry in cooling and heating}
In many realistic systems heating is faster than cooling which has recently been studied in detail \cite{Lapolla,Hasegawa_2021,Godec_2024,Lasanta}. On a phenomenological level this can be modelled with a nonlinear invertible {\it cooling function} $f(T)$

\begin{equation}
    \label{eq:extension}
    {\dot T_\text{int} (t) } = -\kappa   f( T_\text{int} (t) – T_\text{ext}(t) )
\end{equation}
with $f(T)=T$ for small arguments $T$. One possibility is to model the asymmetry by using a $\tanh$-function function 
\begin{equation}
    \label{eq:aaa}
    f(T) = T( 1- v \tanh (T/\sigma_T))
\end{equation}
with a relative prefactor $0<v<1$ and a temperature width $\sigma_T$ such that the effective cooling rate gets smaller than the effective heating rate. In the limit of very small temperature widths $\sigma_T \to 0$ the function $\tanh (T/\sigma_T) = \Theta (T)$ in \eqref{eq:aaa} reduces to a step function.
Since $f(T)$ is monotonic in its argument $T$, it is invertible and the unique inversion is given by 
\begin{equation}
    \label{eq:4aaa}
    T_\text{ext} (t)  = T_\text{int}(t)  -  f^{-1} ( – \dot T_\text{int}(t) /\kappa ).
\end{equation}

\begin{figure}[htbp]
    \centering
    \includegraphics[width=0.99\linewidth]{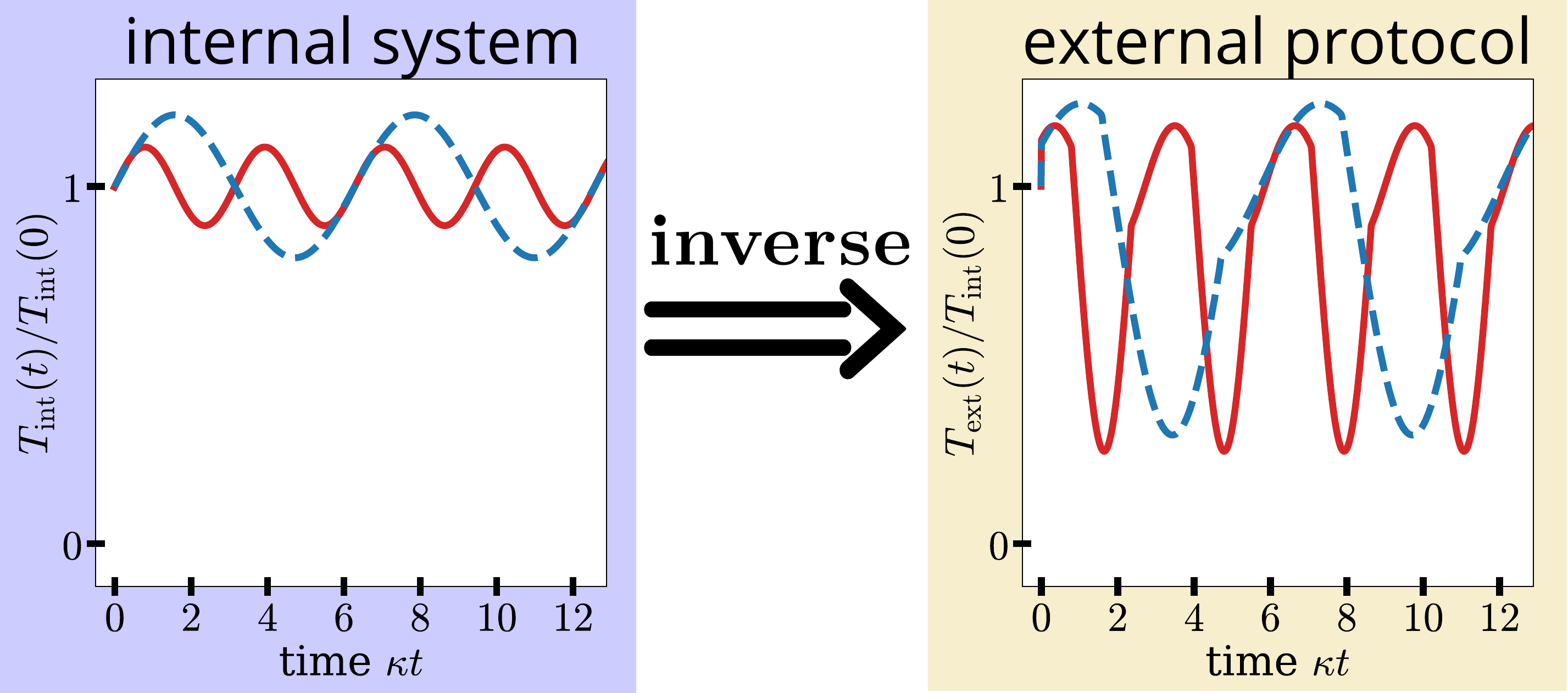}
    \caption{Same as Figure~\ref{fig:1}B b) but now for asymmetric cooling and heating. The time unit is $1/\kappa$ and the temperature unit is $T_\text{int}(0)$. Eqn.\ \eqref{eq:aaa} is used with parameters $v=0.7$ and $\sigma_T = 0.01 T_\text{int}(0)$.}
    \label{fig:6}
\end{figure}

Explicit results for two different sinusoidal cooling and heating cycles are given in Figure~\ref{fig:6}. To maintain the sinusoidal cooling curve the external temperature is much closer to the prescribed one for system heating (i.e.\ when $T_\text{ext}(t) >T_\text{int}(t)$) than for system cooling (i.e.\ when $T_\text{ext}(t) <T_\text{int}(t)$) since the system temperature gradient is much less coupled to the imposed temperature difference $T_\text{ext}(t)- T_\text{int}(t)$ for cooling.

\subsection{Time delay in cooling}
For a realistic system, there can be time-delay between the action of the external bath onto the system. These delay effects have been described by Santos \cite{Santos_2025} by using the equation
\begin{equation}
    \label{eq:delay1}
    {\dot T}_\text{int} (t) = -\kappa ( T_\text{int} (t - \tau) – T_\text{ext}(t) )
\end{equation}
where $\tau >0$ denotes a delay time. In \eqref{eq:delay1} the driving force for a system temperature change is the difference between the actual system temperature $T_\text{ext}(t)$ and a time-delayed system temperature $T_\text{int} (t - \tau)$. If this is assumed, the inversion is unique and direct. For a given desired $T_\text{int} (t )$ the inverse-engineered protocol is simply given by
\begin{equation}
    \label{eq:delay2}
    T_\text{ext}(t) = T_\text{int} (t - \tau) - \dot T_\text{int} (t)/\kappa.
\end{equation}

However, also opposite delay effects are possible. Suppose it takes a typical time $\tau_2$ such that the imposed bath temperature invades into the system and then induces the cooling there. Then system cooling would be described by
\begin{equation}
    \label{eq:delay3}
    {\dot T}_\text{int} (t) = -\kappa_2 ( T_\text{int} (t ) – T_\text{ext}(t-\tau_2 ) )
\end{equation}
rather than by \eqref{eq:delay1}. Now the backward inversion for Eq. \eqref{eq:delay3} has the unique solution
\begin{equation}
    \label{eq:delay4}
    T_\text{ext}(t) = T_\text{int} (t + \tau_2) - {\dot T}_\text{int} (t+\tau_2)/\kappa_2
\end{equation}
as can be easily seen by a shift in time variables.


\section{Do inverse engineered protocols always exist?}
In general, the answer to this question is no. Let us consider a counter-example based on the fact that temperature needs to be non-negative. In the situations discussed so far, the inversion procedure can be carried out formally but the physical constraint that a temperature should be non-negative  leads to restrictions of the inversion: even if $T_\text{int}(t)$ is non-negative for all times $t$, the engineered protocol $T_\text{ext}(t)$ should also be non-negative for all times $t$ \cite{review_Trizac}. If it is negative for certain times, a physical solution of the engineered protocol does not exist.

\begin{figure}[htbp]
    \centering
    \includegraphics[width=0.55\linewidth]{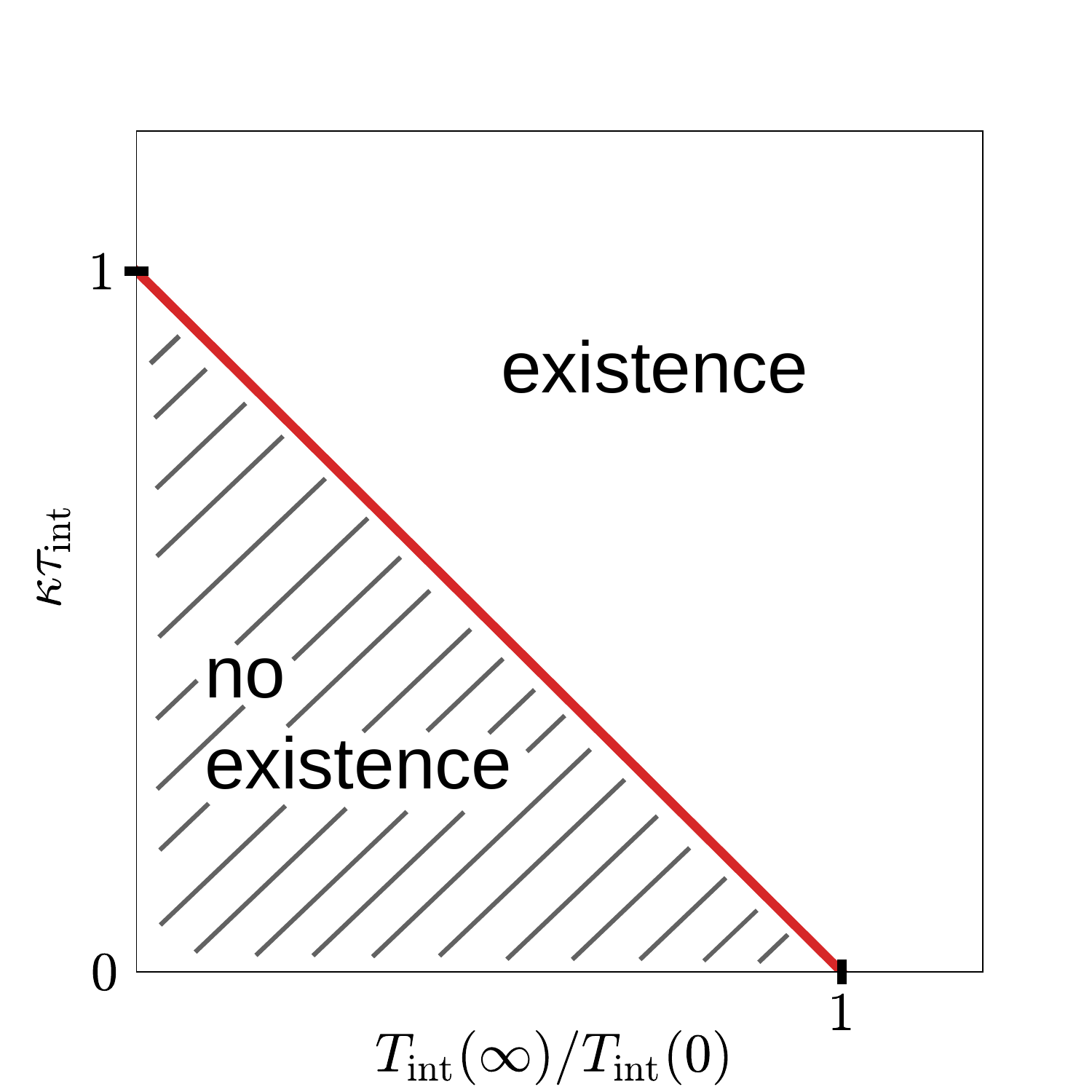}
    \caption{Existence and non-existence of engineered protocols for a prescribed internal cooling curve which decays exponentially in time with a time constant $\tau_\text{int}$ and starts at $T_\text{int}(0)$ for $t=0$ and ends at $T_\text{int}(\infty )$ for $t\to\infty$. For heating, $T_\text{int}(\infty )/T_\text{int}(0)>1$ always a protocol exists.}
    \label{fig:7}
\end{figure}

Consistent with what has been found in earlier studies \cite{Pires}, we exemplify this for the simplest case of Newtonian cooling and a prescribed $T_\text{int}(t)$ that is exponentially decaying for $t>0$ with a decay time $\tau_{int}$ from an initial temperature $T_\text{int}(0)$ to a final temperature $T_\text{int}(\infty )$ as
\begin{equation}
    \label{eq:exist_1}
    T_\text{int}(t) = (T_\text{int}(0) –T_\text{int}(\infty )) \exp (-t/\tau_{int} ) + T_\text{int} (\infty ).
\end{equation}
Clearly the inversion \eqref{eq:6} fulfils the constraint for heating, i.e.\ for $T_\text{int}(\infty )>T_\text{int}(0)$. However,  for cooling $T_\text{int}(\infty )<T_\text{int}(0)$,  \eqref{eq:6} only provides an overall non-negative temperature if the following condition is fulfilled
\begin{equation}
    \label{eq:exist_2}
    \kappa \tau_\text{int} \geq 1-  \frac {T_\text{int}(\infty )} {T_\text{int}(0) }
\end{equation}
which is plotted in Figure~\ref{fig:7}. For small $\tau_\text{int}$, the imposed internal temperature profile is very sharp, hence a significant external undercooling protocol is needed to achieve this sharp decrease such that the protocol falls out of the physical range of positive temperatures.

\section{Are inverse engineered protocols unique?}
Again, the answer to this question is no in general and one can construct a simple but insightful counter-example using a non-monotonic, i.e.\ non-invertible, cooling function $f(T)$.

\begin{figure}[htbp]
    \centering
    \includegraphics[width=0.99\linewidth]{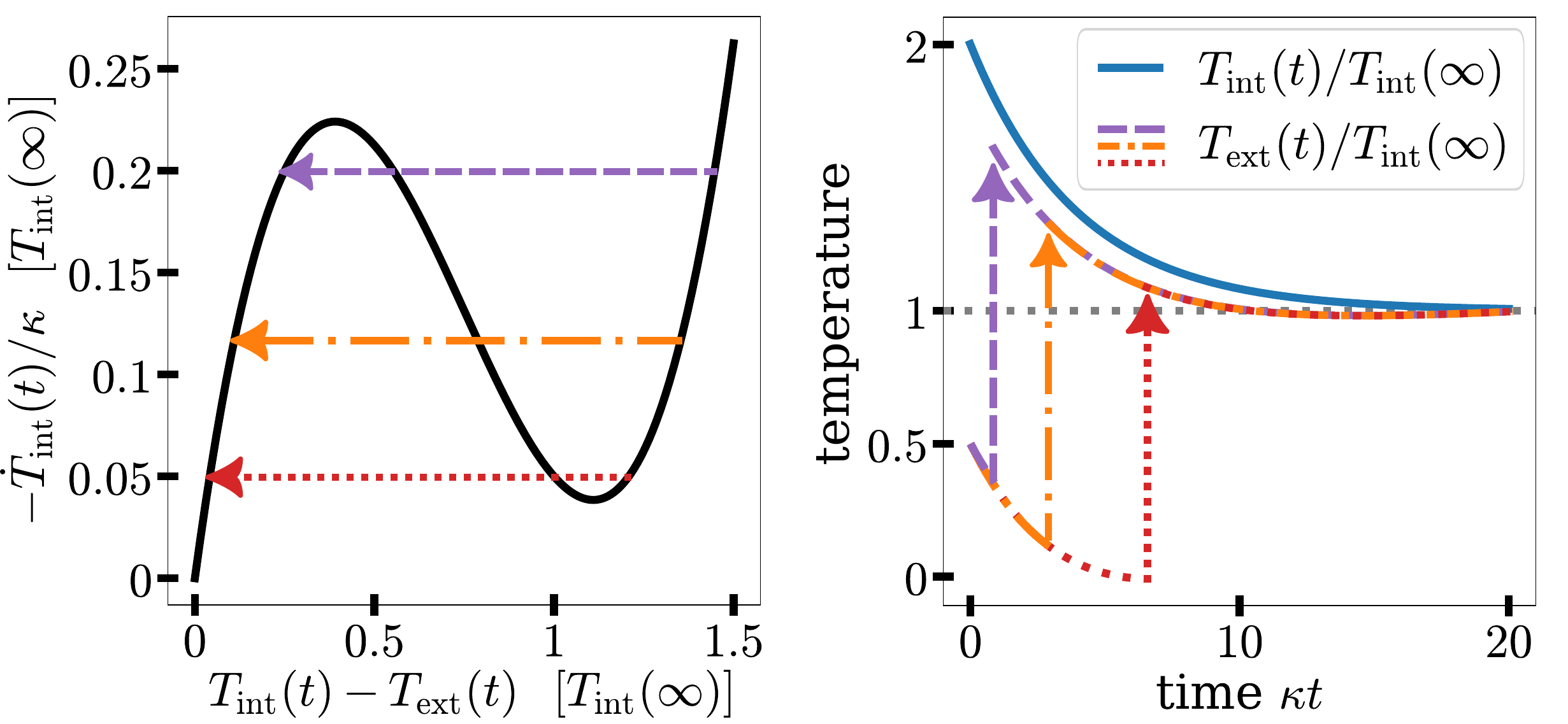}
    \caption{ Left panel: Typical example of a non-monotonic temperature dependence of the  cooling function $f(T_\text{int}(t) - T_\text{ext}(t))$ which is chosen concretely as $f(x)=x^3-2.25x^2+1.3x$ with  $x\equiv (T_\text{int} – T_\text{ext})/T_\text{int}(\infty)$. The exponential cooling curve $T_\text{int}(t)= T_\text{int}(\infty ) + (T_\text{int}(0)- T_\text{int}(\infty ))\exp(-\kappa t/4)$ is prescribed for $t>0$, shown as a blue curve in the right panel. For a given time, the prescribed value of $\dot T_\text{int}(t)/\kappa$ needs to intersect $f(x)$ and therefore multiple solutions are possible. Three of these solutions are shown in different colors, they can be composed together giving rise to a temperature jump in the protocol (colored arrows in both panels). These three different engineered protocols $T_\text{ext}(t)$ lead to the same imposed $T_\text{int}(t)$.}
    \label{fig:8}
\end{figure}

The heat conductivity of a material can be strongly non-monotonic in temperature \cite{Wang}, sometimes called negative differential heat conductivity \cite{negative1,negative2,negative3,Maes}. A classic example is germanium telluride (GeTe)   
\cite{anomalous_heat_conductivity1,anomalous_heat_conductivity2}. Since the cooling rate $\kappa$ is typically dominated by the heat conductivity, non-monotonicities in $\kappa$ can occur which are so strong that they occur also in the full cooling function $f(T)$. We describe these simply within an extension of the Newtonian cooling law to a non-monotonic function $f(T)$ which is not uniquely invertible but unique for small $T$ such that the cooling behaviour gets nonlinear
\begin{equation}
    \label{eq:non-monotonic}
{\dot T_\text{int} (t) } = -\kappa   f( T_\text{int} (t) – T_\text{ext}(t) ).
\end{equation}
Now the inversion is not unique. In general there are several branches for the engineered protocol which can be composed together in different ways but lead to the same prescribed system temperature  $T_\text{int} (t)$. Three of these protocols are shown in Figure \ref{fig:8} which typically exhibit temperature jumps at different times.

An interesting special case occurs when the minimum of the cooling function $f(T)$ coincides with a secondary zero at a temperature difference $T=T_c >0$ apart from the primary zero at $T=0$ such that $f(0)=f(T_c)=0$. This implies that the heat transfer between the external and internal system is blocked at $T=T_c$ corresponding to complete thermal isolation. In this case, again various engineered cooling protocols are possible. If the desired system temperature tends to a final target temperature $T_\text{int}(\infty)$ as $t\to\infty$ (as the blue curve indicated in Figure \ref{fig:8}) a new protocol type arises in which the external temperature is driven to the isolation point $T_\text{int}(\infty) - T_c$ for $t\to \infty$ rather than to $T_\text{int}(\infty)$. Then an arrested thermal transport is maintained between the two final states rather than an established final thermal equilibrium. 


\section{Discussion and Conclusions}
We have computed external cooling protocols in order to engineer a desired system cooling curve. For the traditional Newtonian cooling and the two simple microscopic models considered, there was no Mpemba effect. Then we have extended the phenomenological Newtonian cooling law towards anomalous effects including strong Mpemba, inverse Mpemba and double Mpemba. The effect of overcooling was modelled by an inertial memory term and inverted. Finally an asymmetry between cooling and heating and memory in the system-bath coupling was considered. We then turned to the question of existence and uniqueness of the engineered protocol. Counterexamples are presented which document that existence and uniqueness of the inversion procedure are not guaranteed in general.

The engineered protocols obtained here can be used to realize heat engines which rely on a cyclic sequence of well-defined temperature (and volume) changes. To increase their efficiency it is very useful to be able to perform each step of the engine in a finite time \cite{Blickle,Martinez1,Martinez2} and the engineered protocol can help to optimize this task.

Future problems should address the inverse engineering of cooling protocols in the anomalous case within {\it full microscopic\/} models. It is known that one needs to extend the present considerations significantly either to Brownian particles in non-harmonic potentials (such as bistable  \cite{Kumar_2020,Bechhoefer_2021,Lu_2017,Klich_2019,Malhotra_2024} and asymmetric potentials \cite{Hayakawa_2026}) or to discrete multi-level systems \cite{Melles,discrete_Mpemba}. Then analytical solutions are scarce such that  one has to resort on numerical techniques  for the inversion procedure such as machine learning \cite{Barros,Casert}. For a microscopic model, one has to precisely define system temperature. The harmonic confining potential provides the width of the density distribution as a natural measure for temperature but this is less clear for a bistable potential where approximative distance measures need to be invoked to define system temperature \cite{distance}. This is more clear for an underdamped system such as granular particles which have a well-defined velocity. Then the mean kinetic energy per particle provides a simple and direct measure for system temperature, see e.g.\ \cite{Mandal_Liebchen,wetting}.

Another line of future research concerns active matter \cite{Antonov_2026,Vrugt_2025,Bechinger_2016,Elgeti_2015,Marchetti_2013,Ramaswamy_2010,Bowick_2022} where Mpemba effects can happen as well \cite{Schwarzendahl_2022,Indians}. A promising avenue is an analytical solution for a particle in a harmonic potential \cite{Casert} where the inverted activity protocols \cite{Davis,Olsen2} show analogies to an active jerky harmonic oscillator \cite{jerky}. In this case, compared to passive systems, activity can help to short-cut cooling under certain conditions \cite{Olsen2}.
Finally it would be helpful to combine the engineered protocols with further constraints such as minimal work input (see e.g.\ \cite{Loos}) to achieve a prescribed temperature change as relevant for optimal control theory. 

\section*{Acknowledgements}
I thank Kristian S. Olsen, Maxim Root, Remi Goerlich, Jannis Melles and Udo Seifert for helpful discussions.





%

\end{document}